\documentclass[9pt,a4paper]{article} % 9pt : font size, document class article.cls
\usepackage[hmarginratio=1:1,vmarginratio=1:1,margin=2cm,top=2cm]{geometry}
\usepackage[utf8]{inputenc}
\usepackage[T1]{fontenc}
\usepackage{textcomp}
\usepackage{graphicx}
\usepackage{import}
\usepackage{subcaption}
\usepackage{amssymb}
\usepackage{float}
\usepackage{caption}
\usepackage{svg}
\usepackage{amsmath}
\usepackage{makecell}
\usepackage{textcomp}
\usepackage{setspace}

\usepackage{lineno}
\usepackage{hyperref}
\usepackage{cleveref} 
\usepackage{orcidlink}
\usepackage{authblk}
\usepackage{tabto}
\NumTabs{8}

\usepackage[numbers]{natbib}
\usepackage{tikz}
\usepackage{everypage}
\usepackage{xcolor}

\AddEverypageHook{%
  \begin{tikzpicture}[remember picture, overlay]
    \fill[left color=gray!5,right color=gray!10] 
      ([yshift=0cm]current page.south west) rectangle
      ([xshift=1.2cm,yshift=-0cm]current page.north west);
    \node[rotate=90, anchor=center, text width=\paperheight, align=center, font=\bfseries\small, text=gray]
      at ([xshift=0.6cm]current page.west|-current page.center)
      {This version has been peer-reviewed and accepted for publication by Elsevier}; % Accepted Manuscript (Postprint) - Peer-reviewed version accepted for publication by Elsevier
  \end{tikzpicture}%
}

\title{\centering Towards a better understanding of abdominal wall biomechanics: in vivo relationship between dynamic intra-abdominal pressure and magnetic resonance imaging measurements}

\author[1,2]{Victoria Joppin \orcidlink{0000-0003-2789-6647}}
\author[1]{Arthur Jourdan \orcidlink{0000-0001-8656-4066}}
\author[2]{David Bendahan \orcidlink{0000-0002-1502-0958}}
\author[1]{Andréa Soucasse \orcidlink{0000-0002-5207-1295}}
\author[2,3]{Maxime Guye \orcidlink{0000-0002-4435-2257}}
\author[1]{Catherine Masson \orcidlink{0000-0003-3578-9067}}
\author[1,4]{Thierry Bege \orcidlink{0000-0002-0775-3035}}

\affil[1]{\small Univ Gustave Eiffel, Aix-Marseille Univ, LBA, F-13016 Marseille, France}
\affil[2]{\small Aix Marseille Univ, CNRS, CRMBM UMR 7339, Marseille France}
\affil[3]{\small Assistance Publique-Hopitaux de Marseille, Hôpital Universitaire Timone, CEMEREM, Marseille, France}
\affil[4]{\small Department of General Surgery, Aix-Marseille Univ, North Hospital, APHM, Marseille, France}

\newcommand{\maxIAPbreathing}{11.20}
\newcommand{\maxIAPcough}{57.52}
\newcommand{\maxIAPValsalva}{45.96}
\newcommand{\stdIAPbreathing}{11.15}
\newcommand{\stdIAPcough}{30.88}
\newcommand{\stdIAPValsalva}{26.62}

\newcommand{\maxIAPValsalvamales}{59.14}
\newcommand{\maxIAPValsalvafemales}{29.49}

\newcommand{\maxVAbreathing}{48.27}
\newcommand{\maxVAcough}{-18.99}
\newcommand{\maxVAValsalva}{-16.32}
\newcommand{\stdVAbreathing}{20.69}
\newcommand{\stdVAcough}{8.13}
\newcommand{\stdVAValsalva}{8.04}

\newcommand{\maxdispLMbreathing}{8.86}
\newcommand{\maxdispLMcough}{-7.32}
\newcommand{\maxdispLMValsalva}{-6.52}
\newcommand{\stddispLMbreathing}{4.04}
\newcommand{\stddispLMcough}{2.86}
\newcommand{\stddispLMValsalva}{2.99}

\newcommand{\maxdispRAbreathing}{30.16}
\newcommand{\maxdispRAcough}{-2.63}
\newcommand{\maxdispRAValsalva}{-1.14}
\newcommand{\stddispRAbreathing}{12.33}
\newcommand{\stddispRAcough}{6.38}
\newcommand{\stddispRAValsalva}{6.15}

\newcommand{\maxpearsonIAPVAbreathing}{0.90}
\newcommand{\maxpearsonIAPVAcough}{-0.99}
\newcommand{\maxpearsonIAPVAValsalva}{-0.92}
\newcommand{\stdpearsonIAPVAbreathing}{0.09}
\newcommand{\stdpearsonIAPVAcough}{0.02}
\newcommand{\stdpearsonIAPVAValsalva}{0.07}

\newcommand{\compliancebreathing}{7.65}
\newcommand{\compliancecough}{-0.54}
\newcommand{\complianceValsalva}{-0.56}
\newcommand{\stdcompliancebreathing}{3.97}
\newcommand{\stdcompliancecough}{0.38}
\newcommand{\stdcomplianceValsalva}{0.39}

\newcommand{\compliancebreathingmales}{5.79}
\newcommand{\compliancebreathingfemales}{9.73}
\newcommand{\complianceValsalvamales}{-0.34}
\newcommand{\complianceValsalvafemales}{-0.82}

\newcommand{\maxpearsonIAPdispLMbreathing}{0.78}
\newcommand{\maxpearsonIAPdispLMcough}{-0.98}
\newcommand{\maxpearsonIAPdispLMValsalva}{-0.94}
\newcommand{\stdpearsonIAPdispLMbreathing}{0.45}
\newcommand{\stdpearsonIAPdispLMcough}{0.03}
\newcommand{\stdpearsonIAPdispLMValsalva}{0.05}

\newcommand{\maxpearsonIAPdispRAbreathing}{0.92}
\newcommand{\maxpearsonIAPdispRAcough}{-0.16}
\newcommand{\maxpearsonIAPdispRAValsalva}{0.06}
\newcommand{\stdpearsonIAPdispRAbreathing}{0.06}
\newcommand{\stdpearsonIAPdispRAcough}{0.91}
\newcommand{\stdpearsonIAPdispRAValsalva}{0.91}

\newcommand{\Pvaluethreshold}{0.05}
\newcommand{\Pearsonvaluethreshold}{0.468}

\newcommand{\captionone}{
    Typical 2D axial abdominal MRI slice without annotation a) \\ Segmentation masks are indicated in b) and c)
}
\newcommand{\figureMetricsMRI}{
\begin{figure}[H]
\centering
\includegraphics[width=0.7\textwidth]{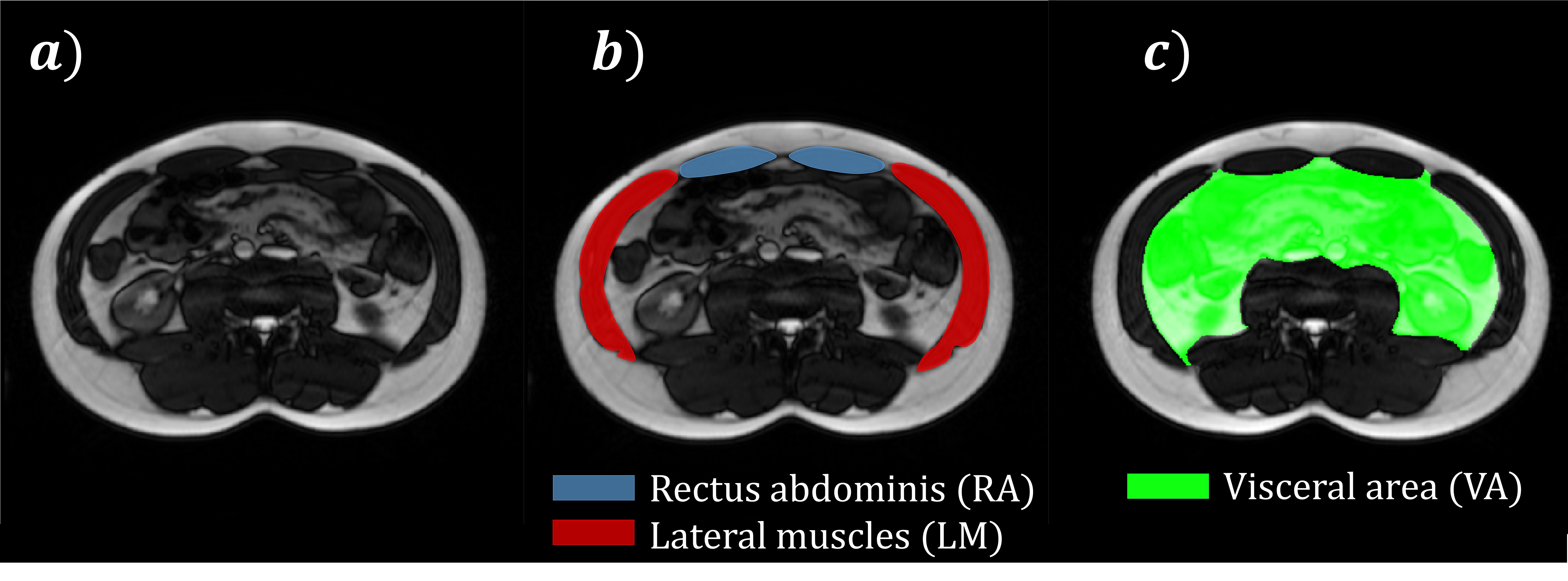}
\captionsetup{justification=centering, format=plain}
\caption[Segmentation of abdominal MRI]{\captionone}
\label{fig:metrics_MRI}
\end{figure}
}

\newcommand{\captiontwo}{
    Metrics derived from MRI \\
a) Abdominal radial distances at rest \\ b) Radial displacement scheme \\ c) Radial displacement of each pixel of the segmentation masks \\ d) Radial displacement averaged among LM and RA segmentation masks
}
\newcommand{\figureRadialDispCsys}{
\begin{figure}[H]
\centering
\includegraphics[width=0.7\textwidth]{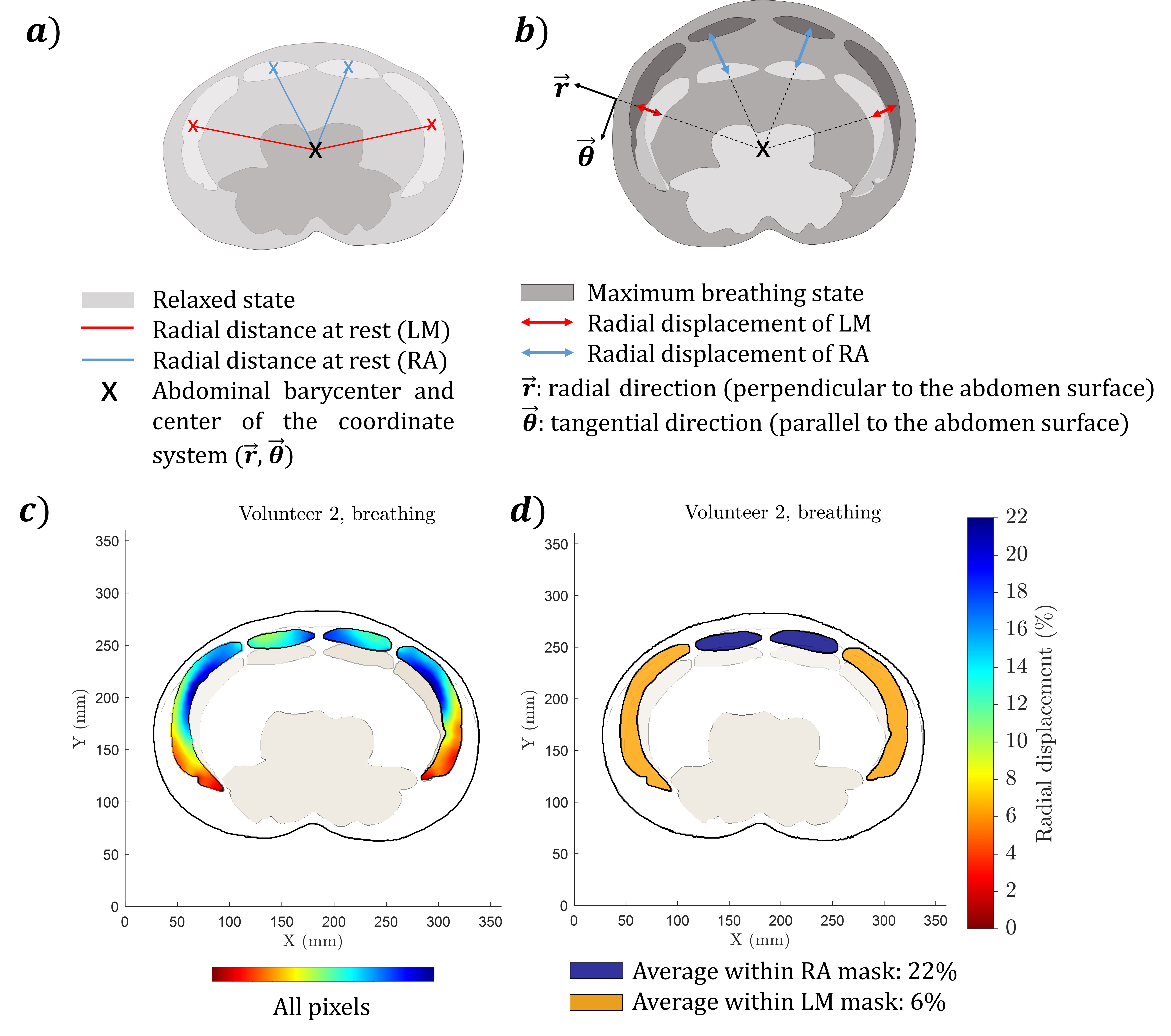}
\captionsetup{justification=centering, format=plain}
\caption[Metrics derived from MRI]{\captiontwo}
\label{fig:radial_disp_csys}
\end{figure}
}

\newcommand{\captionthree}{
    Example of selected timepoints and averaging repetitions during breathing \\ a) Repetitions delimitation (left: visceral area, right: intra-abdominal pressure (IAP)) \\ b) Average cycle (left: visceral area, right: IAP) \\ c) Linear correlation between the average cycles of IAP and visceral area during ascending phase, i.e. from the beginning until the IAP reaches its maximum value
}
\newcommand{\figureAveragingRepetitions}{
\begin{figure}[H]
\centering
\includegraphics[width=0.8\textwidth]{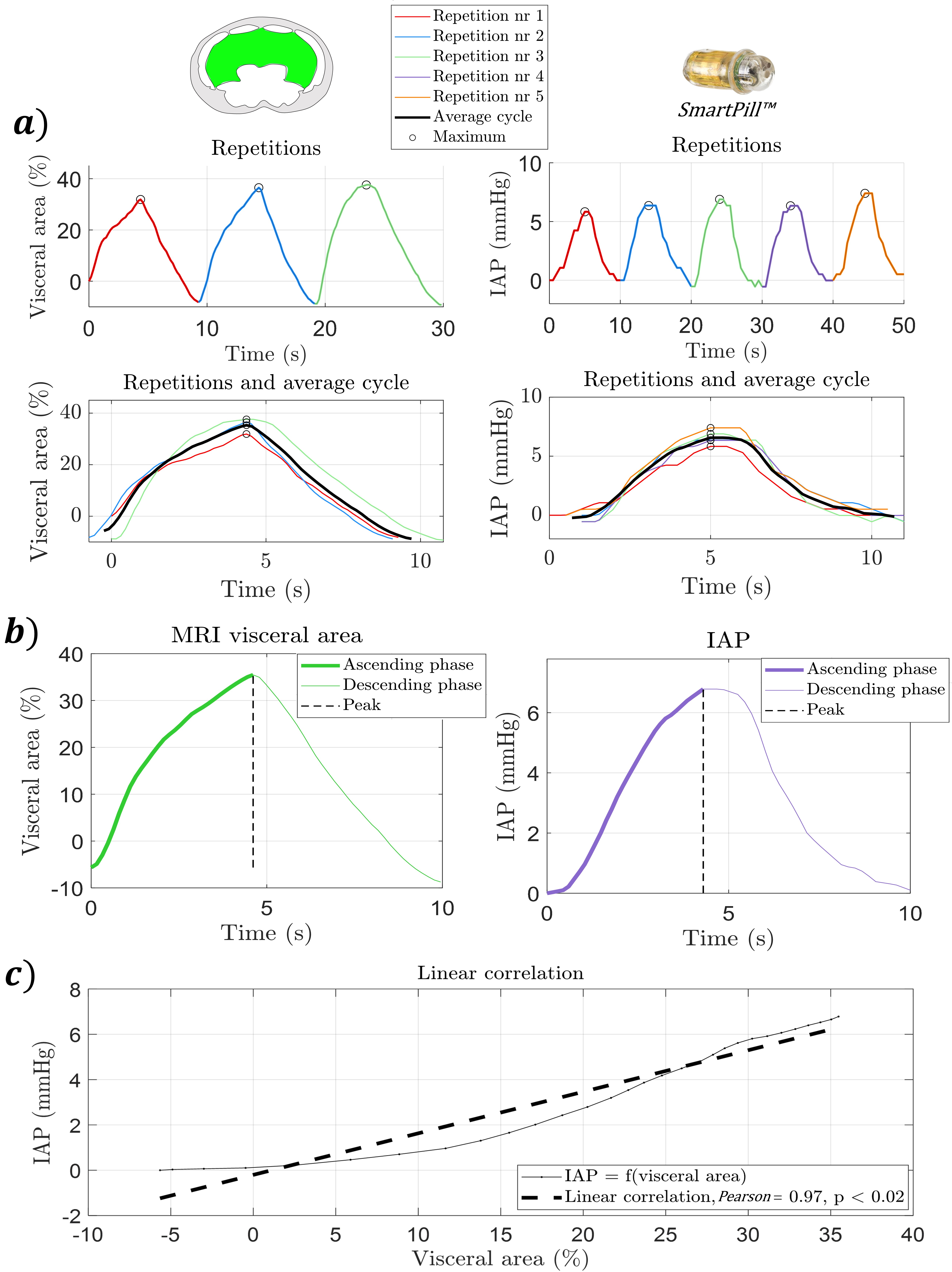}
\captionsetup{justification=centering, format=plain}
\caption[Averaging process of exercises repetitions]{\captionthree}
\label{fig:averaging_repetitions}
\end{figure}
}

\newcommand{\captionfour}{
Temporal changes for different exercises of \\ a) Intra-abdominal pressure (IAP, mmHg) \\ b) Visceral area (\%) \\ c) Radial displacement of LM and RA (\%) \\ Results are presented as mean values $\pm$ one standard deviation $\sigma$
}
\newcommand{\figureMRIandIAPintervolunteers}{
\begin{figure}[H]
\centering
\includegraphics[width=0.75\textwidth]{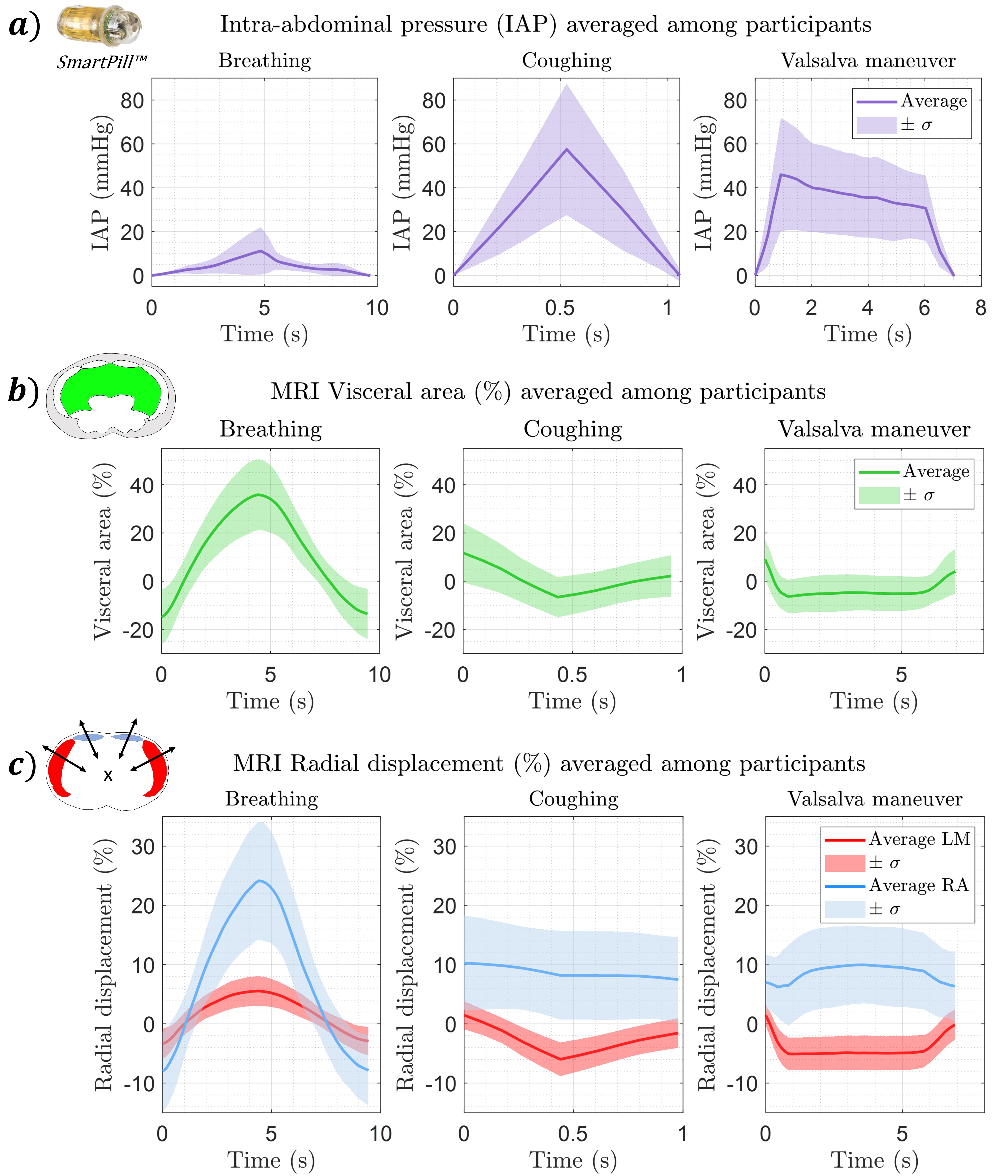}
\captionsetup{justification=centering, format=plain}
\caption[Evolution of pressure and abdominal wall motion]{\captionfour}
\label{fig:MRI_and_IAP_inter_volunteers}
\end{figure}
}

\newcommand{\captionfive}{
Maximum changes in intra-abdominal pressure $\Delta_{max} IAP$, visceral area $\Delta_{max} VA$ and radial displacement of LM and RA for different exercises \\ Results are presented as mean values $\pm$ one standard deviation $\sigma$
}
\newcommand{\tableMaxVariationsIAPMRIVisceralArea}{
\begin{table}[H]
\renewcommand{\arraystretch}{2}
\centering
\begin{center}
\resizebox{\columnwidth}{!}{%
\begin{tabular}{ |c|c|c|c| } 
\hline
\textbf{Exercise} & \makecell{\textbf{Maximum IAP} \\ \textbf{ $\Delta_{max} IAP$ (mmHg)}} & \makecell{\textbf{Maximum visceral area} \\ \textbf{$\Delta_{max} VA$ (\%)}} &
\begin{tabular}{p{3.2cm}|p{3.2cm}}
    \multicolumn{2}{c}{\textbf{Maximum radial displacement (\%)}} \\
    \hline
    \centering \makecell{\textbf{Lateral muscles} \\ \textbf{(LM)}} & \centering \makecell{\textbf{Rectus abdominis} \\\textbf{(RA)}}
\end{tabular} \\
\hline
\textbf{Breathing} & $\maxIAPbreathing$ $\pm$ $\stdIAPbreathing$ & $\maxVAbreathing$ $\pm$ $\stdVAbreathing$ &
\begin{tabular}{p{3.2cm}|p{3.2cm}} 
    \centering $\maxdispLMbreathing$ $\pm$ $\stddispLMbreathing$ & \centering $\maxdispRAbreathing$ $\pm$ $\stddispRAbreathing$
\end{tabular} \\
\hline
\textbf{Coughing} & $\maxIAPcough$ $\pm$ $\stdIAPcough$ & $\maxVAcough$ $\pm$ $\stdVAcough$ &
\begin{tabular}{p{3.2cm}|p{3.2cm}}
    \centering $\maxdispLMcough$ $\pm$ $\stddispLMcough$ & \centering $\maxdispRAcough$ $\pm$ $\stddispRAcough$
\end{tabular} \\
\hline
\textbf{Valsalva} & $\maxIAPValsalva$ $\pm$ $\stdIAPValsalva$ & $\maxVAValsalva$ $\pm$ $\stdVAValsalva$ &
\begin{tabular}{p{3.2cm}|p{3.2cm}}
    \centering $\maxdispLMValsalva$ $\pm$ $\stddispLMValsalva$ & \centering $\maxdispRAValsalva$ $\pm$ $\stddispRAValsalva$
\end{tabular} \\
\hline
\end{tabular}
}
\end{center}
\captionsetup{justification=centering, format=plain}
\caption[Maximum values in pressure and abdominal wall motion]{\captionfive}
\label{tab:max_variations_IAP_MRI_visceral_area}
\end{table}
}

\newcommand{\captionsixe}{
Pearson correlation coefficients (\textit{r}) for the correlations between intra-abdominal pressure (IAP) and visceral area, range [min/max] and coefficient of variation (CV)\\ Results are presented for the different exercises as mean values $\pm$ one standard deviation $\sigma$
}
\newcommand{\tablePearsonIAPVisceralArea}{
\begin{table}[H]
\renewcommand{\arraystretch}{2}
\centering
\begin{center}
\begin{tabular}{ |c|c|c|c| } 
\hline
\textbf{Exercise} & \textbf{\textit{r}} & \textbf{Range [min/max]} & \textbf{CV ($\%$)} \\
\hline
\textbf{Breathing} & $\maxpearsonIAPVAbreathing$ $\pm$ $\stdpearsonIAPVAbreathing$ & [$0.64/1$] & 10 \\
\hline
\textbf{Coughing} & $\maxpearsonIAPVAcough$ $\pm$ $\stdpearsonIAPVAcough$ & [$-1/-0.91$] & 2.02 \\
\hline
\textbf{Valsalva} & $\maxpearsonIAPVAValsalva$ $\pm$ $\stdpearsonIAPVAValsalva$ & [$-1/-0.76$] & 7.61 \\
\hline
\end{tabular}
\end{center}
\captionsetup{justification=centering, format=plain}
\caption[Pearson correlation coefficients between pressure and visceral area]{\captionsixe}
\label{tab:pearson_IAP_visceral_area}
\end{table}
}

\newcommand{\captionseven}{
Intra-abdominal pressure maximum change $\Delta_{max} IAP$ as a function of visceral area maximum change $\Delta_{max} VA$ for all participants and all exercises \\ Each spoke represents one participant (Pn: participant n) \\ Data normalised by the highest value (among all participants) to be scaled from 0 to 1
}
\newcommand{\figureNoLinearCorrIAPMRIVisceralArea}{
\begin{figure}[H]
\centering
\includegraphics[width=0.9\textwidth]{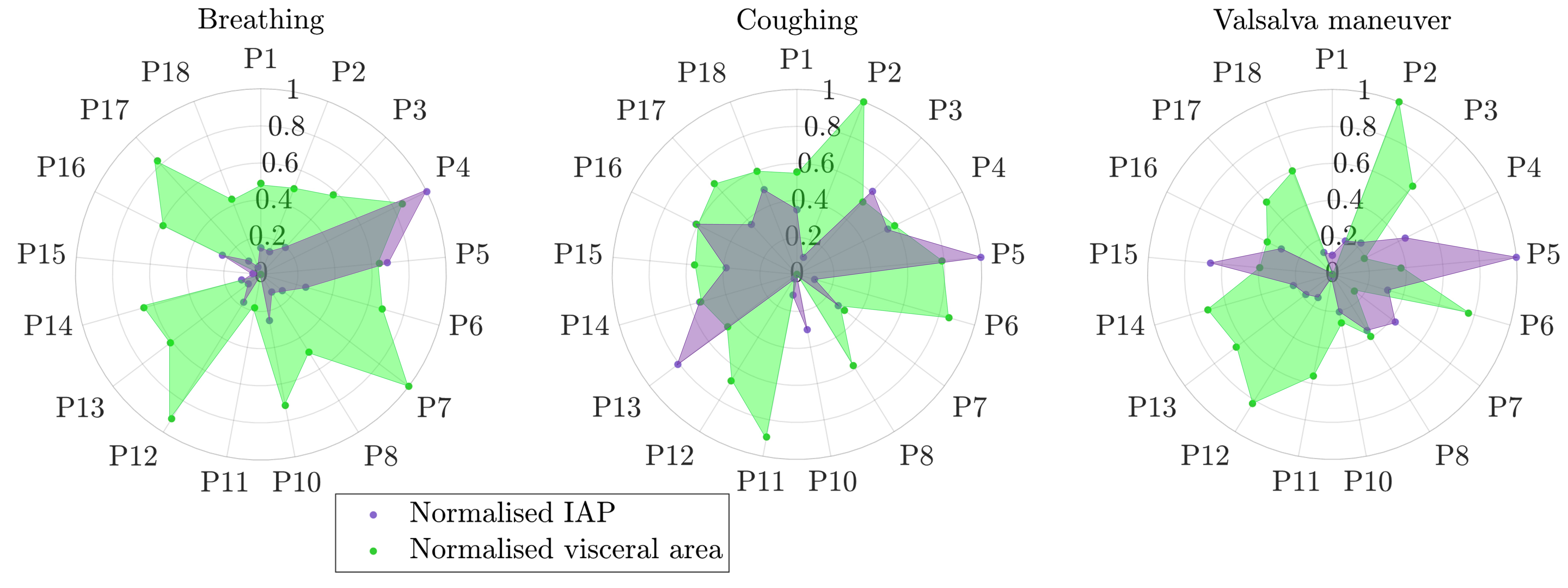}
\captionsetup{justification=centering, format=plain}
\caption[Pressure as a function of visceral area]{\captionseven}
\label{fig:no_linear_corr_IAP_MRI_visceral_area_variation}
\end{figure}
}

\newcommand{\captioneight}{
Pearson correlation coefficients (\textit{r}) for the correlations between intra-abdominal pressure (IAP) and radial displacement of lateral muscles (LM) and rectus abdominis (RA), range [min/max] and coefficient of variation (CV) \\ Results are presented for different exercises as mean values $\pm$ one standard deviation $\sigma$ \\ \textsuperscript{(¤)} indicates that the correlation is not significant for all participants
}
\newcommand{\tablePearsonIAPRadialDispLMRA}{
\begin{table}[H]
\renewcommand{\arraystretch}{2}
\centering
\begin{center}
\resizebox{\columnwidth}{!}{%
\begin{tabular}{ |c|c|c| } 
\hline
\textbf{Exercise} &
\begin{tabular}{p{2.1cm}|p{2.3cm}|p{1.1cm}}
    \multicolumn{3}{c}{\textbf{Lateral muscles (LM)}} \\
    \hline
    \centering \textbf{\textit{r}} & \centering \textbf{Range [min/max]} & \centering \textbf{CV ($\%$)}
\end{tabular} &
\begin{tabular}{p{2.5cm}|p{2.3cm}|p{1.1cm}}
    \multicolumn{3}{c}{\textbf{Rectus abdominis (RA)}} \\
    \hline
    \centering \textbf{\textit{r}} & \centering \textbf{Range [min/max]} & \centering \textbf{CV ($\%$)}
\end{tabular} \\
\hline
\textbf{Breathing} &
\begin{tabular}{p{2.1cm}|p{2.3cm}|p{1.1cm}} 
    \centering $\maxpearsonIAPdispLMbreathing$ $\pm$ $\stdpearsonIAPdispLMbreathing$ & \centering $[-0.94/0.99$] & \centering 57.69
\end{tabular} &
\begin{tabular}{p{2.5cm}|p{2.3cm}|p{1.1cm}} 
    \centering $\maxpearsonIAPdispRAbreathing$ $\pm$ $\stdpearsonIAPdispRAbreathing$ & \centering [$0.79/0.99$] & \centering 6.52
\end{tabular} \\
\hline
\textbf{Coughing} & 
\begin{tabular}{p{2.1cm}|p{2.3cm}|p{1.1cm}} 
    \centering $\maxpearsonIAPdispLMcough$ $\pm$ $\stdpearsonIAPdispLMcough$ & \centering [$-1/-0.86$] & \centering 0.03
\end{tabular} &
\begin{tabular}{p{2.5cm}|p{2.3cm}|p{1.1cm}} 
    \centering $\maxpearsonIAPdispRAcough$\textsuperscript{(¤)} $\pm$ $\stdpearsonIAPdispRAcough$ & \centering [$-0.99/1$] & \centering 600
\end{tabular} \\
\hline
\textbf{Valsalva} & 
\begin{tabular}{p{2.1cm}|p{2.3cm}|p{1.1cm}} 
    \centering $\maxpearsonIAPdispLMValsalva$ $\pm$ $\stdpearsonIAPdispLMValsalva$ & \centering [$-0.99/-0.81$] & \centering 5.32
\end{tabular} &
\begin{tabular}{p{2.4cm}|p{2.3cm}|p{1.1cm}} 
    \centering $\maxpearsonIAPdispRAValsalva$\textsuperscript{(¤)} $\pm$ $\stdpearsonIAPdispRAValsalva$ & \centering [$-0.99/1$] & \centering 1517
\end{tabular} \\
\hline
\end{tabular}
}
\end{center}
\captionsetup{justification=centering, format=plain}
\caption[Pearson correlation coefficients between pressure and muscle displacement]{\captioneight}
\label{tab:pearson_IAP_radial_disp_variation_LM_RA}
\end{table}
}

\begin{document}

\date{Publication date: September 18th, 2024}

\maketitle

\noindent \textbf{Corresponding author}

Victoria Joppin \orcidlink{0000-0003-2789-6647} \tab victoria.joppin@proton.me

Laboratoire de Biomécanique Appliquée, UMRT24 Université Gustave Eiffel - Aix Marseille Université

Faculté des Sciences Médicales et Paramédicales - Secteur Nord

51 Boulevard Pierre Dramard, F-13016 Marseille, France

\vspace{0.5cm}

\noindent \textbf{Authors}

Arthur Jourdan \orcidlink{0000-0001-8656-4066} \tab jourdan.arthur@hotmail.fr

David Bendahan \orcidlink{0000-0002-1502-0958} \tab david.bendahan@univ-amu.fr

Andréa Soucasse \orcidlink{0000-0002-5207-1295} \tab andrea.soucasse@hotmail.fr

Maxime Guye \orcidlink{0000-0002-4435-2257} \tab maxime.guye@ap-hm.fr

Catherine Masson \orcidlink{0000-0003-3578-9067} \tab catherine.masson@univ-eiffel.fr

Thierry Bege \orcidlink{0000-0002-0775-3035} \tab \tab thierry.bege@ap-hm.fr

\vspace{0.5cm}

\section*{Highlights}

\begin{itemize}
    \item Abdominal wall pattern by coupling magnetic resonance image and pressure sensor % 80 characters
    \item Correlation study of muscle motion, visceral area and intra-abdominal pressure change % 85 characters
    \item Lateral muscles were correlated with pressure for active exercise (cough, Valsalva) % 83 characters
    \item Rectus muscles were correlated with pressure for passive exercise (breathing) % 77 characters
    \item Highlights the impact of individual mechanical properties on physiological response % 83 characters
\end{itemize}

\vspace{0.5cm}

\section*{Abstract}
\noindent \textit{Background}: \noindent \textit{In vivo} mechanical behaviour of the abdominal wall has been poorly characterised and important details are missing regarding the occurrence and post-operative recurrence rate of hernias which can be as high as 30$\%$. This study aimed to assess the correlation between abdominal wall displacement and intra-abdominal pressure, as well as abdominal compliance.

\vspace{0.2cm}

\noindent \textit{Methods}: \noindent Eighteen healthy participants performed audio-guided passive (breathing) and active (coughing, Valsalva maneuver) exercises. Axial dynamic changes of abdominal muscles and visceral area were measured using MRI, and intra-abdominal pressure with ingested pressure sensor.

\vspace{0.2cm}

\noindent \textit{Findings}: \noindent Correlations between abdominal wall displacement and intra-abdominal pressure were specific to participant, exercise, and varying between rectus abdominis and lateral muscles. Strong correlations were found between rectus abdominis displacement and intra-abdominal pressure during breathing (\textit{r} = $\maxpearsonIAPdispRAbreathing$ $\pm$ $\stdpearsonIAPdispRAbreathing$), as well as lateral muscles displacement with intra-abdominal pressure during coughing and Valsalva maneuver (\textit{r} = $\maxpearsonIAPdispLMcough$ $\pm$ $\stdpearsonIAPdispLMcough$ and $\maxpearsonIAPdispLMValsalva$ $\pm$ $\stdpearsonIAPdispLMValsalva$ respectively). The abdominal pseudo-compliance varied greatly among participants during muscular contraction, the coefficient of variation reaching up to 70$\%$.

\vspace{0.2cm}

\noindent \textit{Interpretation}: \noindent The combination of intra-abdominal pressure and dynamic MRI measurements enables the identification of participant-specific behaviour pattern. Intra-abdominal pressure and abdominal wall dynamic undergo consistent and predictable interactions. However, this relationship is subject-specific and may not be extrapolated to other individuals. Therefore, both intra-abdominal pressure and abdominal wall motion must be measured in the same participant in order to accurately characterise the abdominal wall behaviour. These results are of great importance for mesh design, surgical decision-making, and personalised healthcare.

\vspace{0.5cm}

%%%%%%%%%%%%
\noindent \textbf{Keywords}

\noindent Abdominal wall muscles; Biomechanics; Dynamic magnetic resonance imaging; Intra-abdominal pressure; Mechanical behaviour % American spelling requested for Keywords

\vspace{0.5cm}

%%%%%%%%%%%%%
\noindent \textbf{Abbreviations}

\noindent IAP: intra-abdominal pressure; LM: lateral muscles; RA: rectus abdominis; MRI: magnetic resonance imaging; \textit{p}: \textit{p}-value; \textit{r}: Pearson correlation coefficient; CV: coefficient of variation; $C_{ab}$: Pseudo abdominal compliance ($cm^2/mmHg$)

\newpage

\section{Introduction}

\noindent The abdominal wall is a multi-layered structure crucial for organ protection, upper body movement and intra-abdominal pressure (IAP) control. However, our understanding of the relationship between IAP and abdominal wall motion is limited, largely due to inadequate investigative methods. Enhanced knowledge of abdominal wall mechanical behaviour could provide valuable insights into physiological and pathological phenomena, such as hernia development. Hernias, characterised by the protrusion of abdominal contents through defects in the musculo-aponeurotic structure, can cause pain and diminish quality of life \cite{smithHealthrelatedQualityLife2022}. Despite advances in surgical treatment, hernia recurrence remains high i.e. 30$\%$ \cite{romainRecurrenceElectiveIncisional2020,flumHaveOutcomesIncisional2003}, likely due to a conflict between the increased IAP leading to an increased abdominal wall loading and the ability of the abdominal wall to cope with this load by moving and deforming.

\vspace{0.2cm}

\noindent Abdominal wall displacements and strains have been assessed \textit{in vivo} mostly externally by image correlation. These studies quantified abdominal motion during exercises \cite{klingeApplicationThreedimensionalStereography1998,todros3DSurfaceImaging2019} and showed that strain depends on anatomical zones \cite{szymczakInvestigationAbdomenSurface2011,smietanskiBiomechanicsFrontAbdominal2012}. However, to fully understand hernia pathophysiology, a detailed examination of the different abdominal wall layers is necessary. Recent research using 2D dynamic MRI has reported specific abdominal muscle displacements \cite{jourdanDynamicMRIQuantificationAbdominal2022}.
While these studies provide valuable data, integrating IAP is critical to fully understand the relationship between abdominal wall motion and pressure, which is essential for understanding both normal and pathological conditions such as hernias.

%%%%%%%%%%%%%%%%%%%%%%%%%%%%%%%%%%%%%

\vspace{0.2cm}

\noindent Most studies on IAP have focused on intraperitoneal hypertension and abdominal compartment syndrome \cite{malbrainResultsInternationalConference2006}, with fewer investigations into its role in hernia formation, although it has been recognised as a potentially important factor \cite{muresanHerniaRecurrenceLong2016}. IAP changes can modulate external abdominal wall surface strain, internal muscular deformations and abdominal volume. Several studies using laparoscopic peritoneal gas insufflation in post-mortem conditions \cite{leruyetDifferencesBiomechanicsAbdominal2020}, anesthetised patients \cite{songMechanicalPropertiesHuman2006} and rabbits \cite{simon-allueMechanicalCharacterizationAbdominal2017} have confirmed the anisotropic behaviour of the abdominal wall, with greater curvature radius \cite{songMechanicalPropertiesHuman2006} and shear modulus \cite{simon-allueMechanicalCharacterizationAbdominal2017} in the longitudinal direction. Song \textit{et al.} showed a near linear correlation between antero-posterior displacement of the abdominal wall and increasing IAP \cite{songMechanicalPropertiesHuman2006}. In post mortem human species, Konerding \textit{et al.} observed a linear correlation between IAP and transverse forces on the linea alba \cite{konerdingMaximumForcesActing2011}. \textit{In vivo} studies also showed that IAP was correlated with electrical muscle activity \cite{misuriVivoUltrasoundAssessment1997,kawabataInteractionBreathingPattern2023}. 

\vspace{0.2cm}

\noindent Abdominal compliance, defined as the ratio of intra-abdominal volume change to IAP change, reflects the abdominal wall’s ability to distend in response to pressure \cite{malbrainRoleAbdominalCompliance2014}. Abdominal compliance has been mainly investigated \textit{in vivo} during laparoscopic peritoneal gas insufflation in porcine models \cite{mcdougallLaparoscopicPneumoperitoneumImpact1994,vlotOptimizingWorkingSpace2013} and humans \cite{mcdougallLaparoscopicPneumoperitoneumImpact1994,abu-rafeaEffectBodyHabitus2006,mulierAbdominalPressureVolume2009,beckerComplianceAbdominalWall2017}. In a porcine model, Vlot \textit{et al.} \cite{vlotOptimizingWorkingSpace2013} reported a 93$\%$ abdominal volume increase when pressure rose from 5 to 10 mmHg. Using CT and surface image correlation, Becker \textit{et al.} \cite{beckerComplianceAbdominalWall2017} reported an exponential relationship between in subcutaneous fat thickness and compliance. In healthy humans, body position (upright or supine) and abdominal binders were shown to influence compliance \cite{calaAbdominalComplianceParasternal1993}. However, compliance during voluntary muscle activation has yet to be studied.

\vspace{0.2cm}

\noindent Despite these advances, gaps remain in our understanding of \textit{in vivo} real-time interactions between IAP and abdominal wall response. Current clinical practices measure IAP invasively via the bladder, but ingestible sensors offer a safer, non-invasive alternative \cite{soucasseAssessmentSmartpillWireless2024,soucasseBetterUnderstandingDaily2022}. Surface imaging techniques are limited to external observations, while ultrasound and elastography only provide localised data. CT scanning, though in-depth, involves radiation. In this context, MRI stands out as a non-invasive, non-radiating alternative that extends the in-depth study of the various muscle groups in the abdominal wall.

\noindent To our knowledge, this is the first study to dynamically assess the \textit{in vivo} relationship between IAP and MRI-based metrics of abdominal wall movement and deformation. Through the combination of dynamic MRI and ingestible pressure sensor measurements in healthy human participants, this study aims to enhance our understanding of the mechanical behaviour of the abdominal wall during various physiological activities that could contribute to hernia development.

\section{Methods}

%%%%%%%%%%%%%%%%%%%%%%%%%%%%
\subsection{Participants}

\noindent Eighteen healthy participants (8 females) volunteered to take part in the study after providing their informed written consent. The average age and BMI were 29.3 $\pm$ 8.1 years old and 22.6 $\pm$ 2.5 $kg/m^2$ respectively.

\noindent This study was approved by the French ethics committee (IDRCB: 2019-A00806–51) and was conducted according to national legislation related to interventional research and the Declaration of Helsinki.

%%%%%%%%%%%%%%%%%%%%%%%%%%%%%%%%%%%%%%%%%%%%%%
\subsection{Protocol}

\noindent Participants were positioned supine in the MRI scanner (MAGNETOM Verio, Siemens Healthineers, Erlangen, Germany) with integrated and flexible body coils on the abdomen. Dynamic (2D+t) MRI captured axial changes every 182 ms during three audio-guided exercises, as previously described \cite{jourdanDynamicMRIQuantificationAbdominal2022}. Breathing (inhale-exhale deeply during 10 s), coughing and the Valsalva maneuver (forceful expiration against closed nostrils and mouth \cite{srivastavValsalvaManeuver2022} during 8 s), chosen as representative of daily activities and clinical exam \cite{jaffeMDCTAbdominalWall2005}, were performed after training.
Breathing was considered passive \cite{campbellVariationsIntraabdominalPressure1953} while coughing and the Valsalva maneuver were active exercises involving abdominal contraction \cite{urquhartAbdominalMuscleRecruitment2005}.

\noindent In a separate session, participants repeated the same exercises after swallowing a SmartPill\textsuperscript{TM} capsule (Medtronic, Minneapolis, MN, USA) to measure IAP ($\pm$ 3.6 mmHg precision) \cite{mccaffreySwallowableCapsuleTechnology2008} with a 500 ms sampling time.

%%%%%%%%%%%%%%%%%%%%%%%%%%%%%%

\subsection{Data processing}

\noindent As illustrated in \autoref{fig:metrics_MRI}-a, muscles of interest were manually delineated in MR images using FSLeyes \cite{smithAdvancesFunctionalStructural2004}, then a semi-automatic segmentation algorithm was used to obtain the segmentation masks at each timestep \cite{ogierIndividualMuscleSegmentation2017}. Segmentation masks included the rectus abdominis (RA) and lateral muscles (LM, including the transverse abdominis, internal and external obliques), as represented in \autoref{fig:metrics_MRI}-b.

%% FIGURE CALLING %%
\figureMetricsMRI

\noindent Radial displacements (\autoref{fig:radial_disp_csys}-b) were computed for each pixel of the segmented muscles \cite{jourdanSemiautomaticQuantificationAbdominal2021} (\autoref{fig:radial_disp_csys}-c), then averaged across the LM and RA muscles (\autoref{fig:radial_disp_csys}-d).

%% FIGURE CALLING %%
\figureRadialDispCsys

\noindent The visceral area (\autoref{fig:metrics_MRI}-c), defined as the region delimited by the internal surface of the abdominal muscles and dorsal structures (quadratus lumborum, psoas major, erector spinae, vertebral disc and body), was computed.

\noindent Radial displacement of LM and RA muscles and visceral area metrics were normalised to their resting values, defined as the endpoint of a calm exhalation without visible contraction, ensuring no variation at rest. For standardisation between participants, these metrics were expressed relative to the resting state: the visceral area was divided by its resting value, and the radial displacements were divided by the radial distance between the barycenter of each muscle group (LM and RA) at resting state and the abdominal barycenter (\autoref{fig:radial_disp_csys}-a). The barycenter of a muscle is defined as the mean position of all pixels that constitute its segmentation mask. Meanwhile, the abdominal barycenter is the average location of all pixels composing the LM segmentation masks.

\noindent As illustrated in \autoref{fig:averaging_repetitions}-a for each exercise, specific timepoints, i.e. onset and peak, were manually identified so as to delimit the different repetitions using MATLAB.

\noindent Then, all repetitions were linearly oversampled to $n$ points, corresponding to the longest repetition across participants. This equates to $n$ = 60, 12 and 54 points for breathing, coughing and the Valsalva maneuver, respectively, due to varying exercise durations. This approach was chosen to achieve a slight upsampling of the shorter repetitions, while avoiding the loss of information associated with downsampling the longer ones. For each participant, the resulting repetitions of $n$ points were averaged to generate a representative average cycle of $n$ points of each exercise (\autoref{fig:averaging_repetitions}-b). As illustrated in \autoref{fig:averaging_repetitions}-a, the maximum values of the each set of repetitions were aligned for signal averaging in order to prevent the peak from being smoothed out.

%% FIGURE CALLING %%
\figureAveragingRepetitions

\vspace{0.1cm}

\noindent The IAP values were exported using the SmartPill\textsuperscript{TM} 3.1 software with the temperature-compensated mode. The IAP repetitions were oversampled to reach $n$ points by linear interpolation and averaged similarly to the MRI data. This process standardised the cycle lengths across repetitions, allowing for the correlation of MRI-based abdominal data and IAP values, despite non-simultaneous measurements. This approach enabled consistent comparison between abdominal motion and IAP across different participants and exercises.

%%%%%%%%%%%%%%%%
\subsection{Correlation intra-abdominal pressure - abdominal wall motion}

\noindent This study investigates the relationship between intra-abdominal pressure (IAP) and MRI-based metrics of abdominal wall movement and deformation.
Pearson correlation coefficients \textit{r} were  calculated for each participant and exercise to evaluate the direction (positive or negative) and strength of correlations between IAP and visceral area, as well as IAP and radial displacement of LM and RA. These correlations were assessed during the ascending phase of each exercise (until maximum or plateau for the Valsalva, as illustrated in \autoref{fig:averaging_repetitions}-c), with statistical significance considered for p-value \textit{P} $\leq$ $\Pvaluethreshold$ and \textit{r} $\geq$ $\Pearsonvaluethreshold$, critical threshold of an acceptable correlation strength \cite{cunninghamAppendixStatisticalTables2013}.

\vspace{0.1cm}

\noindent Abdominal pseudo-compliance was computed as the ratio between the maximum changes in visceral area $\Delta_{max}VA $ (value before normalisation, expressed in $cm^2$) and intra-abdominal pressure $\Delta_{max}IAP $:
\begin{equation*}
    C_{ab} = \frac{\Delta_{max} VA}{\Delta_{max} IAP} (\frac{cm^2}{mmHg})
\end{equation*}

\noindent The maximum change was defined as the difference between the beginning and the end of the ascending phase (\autoref{fig:averaging_repetitions}-b). This metric differs from the conventional compliance using volume given that it is applied to 2D MRI slices.

\noindent Results were presented as average across all participants (n = 18) $\pm$ standard deviation $\sigma$, with their [minimum/maximum] range and coefficient of variation CV, defined as the standard deviation divided by the absolute \textit{r} value: $ CV = \sigma/|\textit{r}|$ ($\%$).

\noindent The correlation between IAP and MRI measurements was further analysed in relation to morphologic parameters, including BMI ($kg/m^2$), waist circumference (mm), radial distance at rest of LM and RA (mm) and muscle mass ratio ($\%$), defined as the area occupied by abdominal muscles divided by the whole abdomen area (delimited by the external surface of the abdomen). The difference according to the sex assigned at birth (female or male) was assessed using a two-sided Wilcoxon-Mann-Whitney test, with a \textit{P} $\leq$ $\Pvaluethreshold$ considered statistically significant.

\section{Results}

%%%%%%%%%%%%%%%%%%%%%%%%%%%%%%%%%%%%%%%%%%%%%%%%%%%%%%%%%%%%%%%
\subsection{Intra-abdominal pressure and MRI time-dependent changes}

\noindent Temporal changes of IAP, visceral area and radial displacement of LM and RA during the different exercises are displayed in \autoref{fig:MRI_and_IAP_inter_volunteers}, with maximum changes summarised in \autoref{tab:max_variations_IAP_MRI_visceral_area}. Radial displacement and visceral area were analysed for 17 participants as one MRI dataset was excluded due to poor quality.

%% FIGURE CALLING %%
\figureMRIandIAPintervolunteers

\noindent During breathing, IAP, visceral area and radial displacement of LM and RA muscles followed a bell shape pattern with ascending and descending phases. Breathing induced a large visceral area increase and an outward/positive radial displacement.

\noindent Coughing also exhibited these two phases, but visceral area and LM displacement initially decreased during contraction, contrasting with IAP's initial rise. Radial displacement of RA was limited ($\maxdispRAcough$ $\pm$ $\stddispRAcough$$\%$) compared to LM ($\maxdispLMcough$ $\pm$ $\stddispLMcough$$\%$), with notable inter-participant variability, RA being pulled in or out according to the participant's coughing execution. The positive RA displacement at the onset was related to a short intake of air before coughing.

\noindent For the Valsalva maneuver, three phases where observed: IAP and RA displacement increased initially, while LM displacement and visceral area decreased. Variability in RA displacement was higher than LM, highlighting again differences in execution among participants (standard deviation of $\stddispRAValsalva$$\%$ for RA compared to $\stddispLMValsalva$$\%$ for LM).

\noindent Activities with muscular activation led to a visceral area decrease accompanied by an inward, negative displacement of LM and a higher $\Delta_{max} IAP$. IAP during coughing was 11 mmHg higher than during the Valsalva maneuver. For the Valsalva maneuver, a significant correlation between $\Delta_{max} IAP$ and the muscle mass ratio was found (\textit{r} = 0.71, \textit{P} $\leq$ $\Pvaluethreshold$), with males showing higher IAP compared to females ($\maxIAPValsalvamales$ and $\maxIAPValsalvafemales$ mmHg, respectively).

%% FIGURE CALLING %%
\tableMaxVariationsIAPMRIVisceralArea

%%%%%%%%%%%%%%%%%%%%%%%%%%%%%%%%%%%%%%%%%%%%%%%%%%%%%%%
\subsection{Relationship between intra-abdominal pressure and visceral area}

\noindent The relationship between IAP and visceral area changes was assessed in order to provide information related to the dynamic behaviour of the abdominal wall on a global scale, i.e. without distinction between LM and RA muscles. The mean \textit{r} values are presented in \autoref{tab:pearson_IAP_visceral_area}.

%% FIGURE CALLING %%
\tablePearsonIAPVisceralArea

\noindent The correlation between IAP and visceral area changes was significant (\textit{P} $\leq$ $\Pvaluethreshold$ and $|\textit{r}|$ $\geq$ $\Pearsonvaluethreshold$) and very subject-specific for each participant and each exercise. For coughing and the Valsalva maneuver, \textit{r} values were negative indicating that IAP was increasing while the visceral area was decreasing. The correlation was strong, with an absolute mean \textit{r} value larger than 0.9 for the whole set of exercises.

\noindent \autoref{fig:no_linear_corr_IAP_MRI_visceral_area_variation} illustrates the maximum changes of intra-abdominal pressure $\Delta_{max} IAP$ and visceral area $\Delta_{max} VA$ for all the participants and the three exercises. Both variables were normalised by their highest values among the 18 participants to be scaled from 0 to 1. The largest IAP change did not always correspond to the largest visceral area change. For example, participant 7 had large area changes with moderate IAP during breathing, whereas participant 5 showed the opposite during the Valsalva maneuver. No significant correlation was found between participants between maximum changes in IAP and visceral area, indicating that their relationship is participant-specific and cannot be generalised to the whole cohort.

%% FIGURE CALLING %%
\figureNoLinearCorrIAPMRIVisceralArea

%%%%%%%%%%%%%%%%%%%

\vspace{0.2cm}

\noindent The mean abdominal pseudo-compliances$\ C_{ab}$ were $\compliancebreathing$ $\pm$ $\stdcompliancebreathing$, $\compliancecough$ $\pm$ $\stdcompliancecough$ and $\complianceValsalva$ $\pm$ $\stdcomplianceValsalva$ $cm^2/mmHg$ for breathing, coughing and the Valsalva maneuver, respectively. The corresponding [minimum/maximum] ranges were [$1.78/13.68$], [$-1.63/-0.13$] and [$-1.40/-0.12$] $cm^2/mmHg$, and coefficients of variation CV of 51.90, 70.37 and 69.64$\%$.

\noindent Participants with high pseudo-compliance during breathing did not necessarily have high pseudo-compliance during muscle contraction (coughing and the Valsalva maneuver). A significant difference in pseudo-compliance was found between males and females for breathing and the Valsalva maneuver exercises. During breathing, females had higher pseudo-compliance than males ($\compliancebreathingfemales$ versus $\compliancebreathingmales$ $cm^2/mmHg$ respectively). Inversely, with muscular activation during the Valsalva maneuver, females had lower pseudo-compliance than males ($\complianceValsalvafemales$ versus $\complianceValsalvamales$ $cm^2/mmHg$, respectively).

%%%%%%%%%%%%%%%%%%%%%%%%%%%%%%%%%%%%%%%%%%%%%%%%%%%%%%%%%%%
\subsection{Relationship between intra-abdominal pressure and displacement}

\noindent The correlation between IAP and abdominal muscles displacements was assessed to explore the dynamic behaviour of the abdominal wall at a local level, distinguishing LM and RA muscles. The mean \textit{r} values related to the correlations are presented in \autoref{tab:pearson_IAP_radial_disp_variation_LM_RA}.

%% FIGURE CALLING %%
\tablePearsonIAPRadialDispLMRA

\noindent A significant correlation was found between IAP and radial displacement of LM for all participants and all exercises. For RA, all participants had a significant correlation during breathing. There were 5 participants who did not have a significant correlation during coughing and 2 participants during the Valsalva maneuver with a \textit{P} $\geq$ $\Pvaluethreshold$.

\noindent The correlation between IAP and radial displacement varied depending on the muscle group (LM or RA) and the type of exercise performed (breathing, coughing or the Valsalva maneuver). IAP and radial displacement of RA were strongly correlated (\textit{r} = $\maxpearsonIAPdispRAbreathing$ $\pm$ $\stdpearsonIAPdispRAbreathing$) during breathing whereas very weak correlations were observed for coughing (\textit{r} = $\maxpearsonIAPdispRAcough$ $\pm$ $\stdpearsonIAPdispRAcough$) and the Valsalva maneuver (\textit{r} = $\maxpearsonIAPdispRAValsalva$ $\pm$ $\stdpearsonIAPdispRAValsalva$). On the contrary, the correlation between IAP and radial displacement of LM was better and had a lower standard deviation during active exercises (\textit{r} = $\maxpearsonIAPdispLMcough$ $\pm$ $\stdpearsonIAPdispLMcough$ and $\maxpearsonIAPdispLMValsalva$ $\pm$ $\stdpearsonIAPdispLMValsalva$ respectively for coughing and the Valsalva maneuver) than breathing (\textit{r} = $\maxpearsonIAPdispLMbreathing$ $\pm$ $\stdpearsonIAPdispLMbreathing$). As for visceral area, no significant linear correlation was found between participants, between maximum changes in IAP and radial displacement of LM or RA. The relationship between the IAP and radial displacement is very different among the participants. 

\section{Discussion}

\noindent This work is the first study to investigate the correlation between IAP and MRI-based abdominal wall metrics. The present results provide valuable insights regarding the \textit{in vivo} relationship between IAP and the structural and functional characteristics of the abdominal wall. This relationship was studied on both global (through the quantification of visceral area) and local scales (though radial displacement of LM and RA muscles), during passive (breathing) and active (coughing, Valsalva maneuver) exercises.

\vspace{0.2cm}

%% Temporal evolution of metrics, pattern
\noindent The IAP, visceral area and muscular radial displacement followed a specific pattern for each exercise. During coughing and Valsalva exercises, the recruitment of abdominal muscles led to a compression of the viscera, thus a decrease in visceral area ($\maxVAcough$ $\pm$ $\stdVAcough$ and $\maxVAValsalva$ $\pm$ $\stdVAValsalva$$\%$ for coughing and the Valsalva maneuver respectively, as compared to +$\maxVAbreathing$ $\pm$ $\stdVAbreathing$$\%$ for breathing). The slight differences in IAP, visceral area and radial displacement between coughing and the Valsalva maneuver may be related to the faster IAP increase during coughing reported by Soucasse \textit{et al.} \cite{soucasseBetterUnderstandingDaily2022}, thereby further supporting that the loading rate influences the behaviour of the abdominal wall tissue \cite{benabdelounisEffectTwoLoading2013}.

\vspace{0.2cm}

%% IAP - visceral area
\noindent The present results allowed to assess the correlation between IAP and visceral area. During passive exercise such as breathing, this linear and positive correlation was in agreement with the results reported \textit{in vivo} by Song \textit{et al.} \cite{songMechanicalPropertiesHuman2006}, showing that the displacement of the abdominal outer surface evolved almost linearly with the insufflated pressure. As a measure of \textit{in vivo} abdominal compliance, Papavramidis \textit{et al.} also showed a linear correlation between abdominal pressure and ascitic fluid drained volume in patients \cite{papavramidisAbdominalComplianceLinearity2011}.

%% Compliance
\noindent Our results showed that abdominal pseudo-compliance based on area rather than volume, was positive and larger during breathing ($\compliancebreathing$ $\pm$ $\stdcompliancebreathing$ for breathing versus $\compliancecough$ $\pm$ $\stdcompliancecough$ and $\complianceValsalva$ $\pm$ $\stdcomplianceValsalva$ for coughing and the Valsalva maneuver respectively), with the abdominal wall shape being markedly stretched without being contracted, leading to a slight IAP increase but an important visceral area increase \cite{campbellVariationsIntraabdominalPressure1953}.

\noindent On the contrary, active muscular exercises induced a visceral area decrease accompanied by high IAP increase, thereby resulting in a negative and smaller pseudo-compliance. In most biological tissues (veins, lungs), increased pressure leads to tissue expansion and positive compliance. However in this case, a decrease in visceral area results in negative pseudo-compliance, likely due to muscle activation (e.g., LM/RA). A review article also reported that abdominal muscles decreased the abdominal compliance \cite{malbrainRoleAbdominalCompliance2014}. During the Valsalva maneuver, males had a significantly higher IAP than females ($\maxIAPValsalvamales$ and $\maxIAPValsalvafemales$ mmHg, respectively), and a higher pseudo-compliance ($\complianceValsalvamales$ versus $\complianceValsalvafemales$ $cm^2/mmHg$ respectively). While previous studies found no correlation between IAP and gender \cite{cobbNormalIntraabdominalPressure2005,sanchezWhatNormalIntraabdominal2001}, males generally have a higher muscle mass ratio than females \cite{jourdanAbdominalWallMorphometric2020,rankinAbdominalMuscleSize2006}, as seen in this cohort (16.25 versus 13$\%$ for males and females respectively, \textit{P} $\leq$ $\Pvaluethreshold$). The muscle mass ratio being correlated with IAP changes during Valsalva, males can generate a higher contraction. During breathing, females had a higher pseudo-compliance than males ($\compliancebreathingfemales$ versus $\compliancebreathingmales$ $cm^2/mmHg$ respectively), in line with the fact that females had a higher abdominal distention capacity \cite{jungeElasticityAnteriorAbdominal2001}.

\vspace{0.2cm}

%% IAP - Radial displacement
\noindent Our findings indicate that correlations between IAP and radial displacement differ according to the type of exercise and the muscular group. Breathing resulted in a stronger correlation between IAP and the radial displacement of RA (\textit{r} = $\maxpearsonIAPdispRAbreathing$ for RA versus $\maxpearsonIAPdispLMbreathing$ for LM). Conversely, during active exercise, the LM appeared more involved than RA with a stronger correlation with IAP (for LM, \textit{r} = $\maxpearsonIAPdispLMcough$ and $\maxpearsonIAPdispLMValsalva$ for coughing and the Valsalva maneuver respectively compared $\maxpearsonIAPdispRAcough$ and $\maxpearsonIAPdispRAValsalva$ for RA). The correlation between IAP and radial displacement of RA was not significant for all participants during coughing and Valsalva, with a \textit{P} $\leq$ $\Pvaluethreshold$. This may be due to the fact that these muscles are less involved than LM during contraction, and these specific participants moved their RA minimally during these exercises. Also, the p-value \textit{P} is affected by the number of datapoints, and these exercises are shorter than breathing. Misuri \textit{et al.} \cite{misuriVivoUltrasoundAssessment1997} also reported that IAP was linearly related to the thickness of the transverse muscle, being part of the LM considered in this study. Urquhart \textit{et al.} also showed that LM were more recruited than RA during voluntary exercises \cite{urquhartAbdominalMuscleRecruitment2005}.

\vspace{0.2cm}

\noindent At the intra-individual level, a robust correlation has been identified between IAP and abdominal wall motion, suggesting that these variables are tightly linked by participant-specific relationship. Each participant exhibits a unique pseudo-compliance for each exercise, which can be modelled as IAP = $f$(abdominal wall motion). Once this function $f$ has been identified for a participant, one variable can be predicted from the other. This relationship encompasses factors like anatomy, abdominal wall behaviour, and exercise execution. Regardless of the exercise, the correlation between changes in IAP and visceral area was strong, with \textit{r} values above 0.9. The correlation between IAP and the radial displacement of LM during active exercises and the RA during passive ones was also strong for each participant.

\vspace{0.2cm}

\noindent Nevertheless, at the inter-individual level, the results showed a high variability in IAP and abdominal wall motion. No consistent pattern was identified in which a high IAP was systematically accompanied by a large visceral area or radial displacement of LM and RA. In other words, a participant with high pseudo-compliance may experience substantial change in visceral area with minimal IAP increases. The compliance parameter accounts for a considerable range of combinations between pressure values and visceral areas. The variability in the above-mentioned function $f$ between participants reflects anatomical and mechanical differences. Previous studies have highlighted these inter-individual differences in abdominal anatomy \cite{jourdanAbdominalWallMorphometric2020} and mechanical properties both \textit{ex vivo} \cite{jungeElasticityAnteriorAbdominal2001,martinsMechanicalCharacterizationConstitutive2012} and \textit{in vivo} \cite{johnCanActivityExternal2007}. Szepietowska \textit{et al.} \cite{szepietowskaFullfieldVivoExperimental2023} also reported recently an apparent lack of correlation between the maximum IAP values and abdominal strains between participants, further reinforcing the need for individualised measurements of both IAP and abdominal wall motion to accurately assess mechanical behaviour.

\vspace{0.2cm}

%%%%%%%%%%%%%%%%%%%%%%% LIMITATIONS %%%%%%%%%%%%%%%%%%%%

\noindent Several limitations should be acknowledged in this study. Dynamic MRI was recorded in a single axial plane, perpendicular to the RA muscle fibres. A 3D dynamic analysis would lengthen the time resolution to 1s instead of 182 ms, insufficient to capture contraction. Also, it has been documented that abdominal muscles movement in the cranio-caudal direction is minimal \cite{jourdanDynamicMRIQuantificationAbdominal2022}. Another limitation is the non-simultaneous IAP and MRI measurements, as the SmartPill\textsuperscript{TM} sensor, with a different sampling frequency, is not MRI compatible. To the best of our knowledge, there are no MRI-compatible IAP sensor. To ensure exercise repeatability and accurate motion capture, participants were trained under similar conditions. Before each MRI and SmartPill\textsuperscript{TM} session, the exercises were audio-guided and repeated multiple times. Oversampling shorter repetitions to match the longest repetition may introduce bias. Secondly, the hypothesis that maximum pressure and maximum abdominal wall displacement occur simultaneously is supported by \textit{in vivo} animal study \cite{vlotOptimizingWorkingSpace2013} where the antero-posterior distance and abdominal volume reached their maximum values when pressure reached its maximum. Human studies combining pressure and trunk motion \cite{hodgesChangesIntraabdominalPressure2000} are limited.

\vspace{0.2cm}

%%%%%%%%%%%%%%%%%%%%%%% CLINICAL APPLICATIONS %%%%%%%%%%%%%%%%%%%%

\noindent This study on healthy participants is a prerequisite for assessing patients with abdominal wall pathologies, such as hernias. Extending these investigations to such groups would provide insights into mechanical alterations related to these conditions. It is reasonable to assume that the heterogeneity in mechanical characteristics amongst participants may impact the outcomes (pain, recurrence) of hernia repair procedures. A patient-based study could lead to the development of personalised mesh solutions, surgical interventions, thereby enhancing the effectiveness of hernia repair and minimising associated complications through the early diagnosis of abdominal wall changes.

\section{Conclusion}

\noindent This work is the first to investigate the relationship between intra-abdominal pressure and dynamic MRI-based abdominal wall metrics through continuous temporal \textit{in vivo} measurements in healthy human subjects. The present results illustrate the interest of combining these two measurements to assess dynamically the mechanical behaviour of the abdominal wall. There was a strong correlation between abdominal wall motion and IAP, which was dependent on the participant. The corresponding results provide interesting information related to abdominal wall dynamics and could be used for the evaluation of patients with abdominal wall hernias.

%%%%%%%%%%%%%%%%%%%%%%%%%%%%%%%%%
\section*{Citation Diversity Statement}

\noindent Recent work in several fields of science has identified a bias in citation practices such that papers from women and other minority scholars are under-cited relative to the number of such papers in the field \cite{mitchell2013gendered,dion2018gendered,caplar2017quantitative, maliniak2013gender, Dworkin2020.01.03.894378, bertolero2021racial, wang2021gendered, chatterjee2021gender, fulvio2021imbalance}. Here we sought to proactively consider choosing references that reflect the diversity of the field in thought, form of contribution, gender, race, ethnicity, and other factors.

\vspace{0.3cm}

\noindent First, we obtained the predicted gender of the first and last author of each reference by using databases that store the probability of a first name being carried by a woman \cite{Dworkin2020.01.03.894378,zhou_dale_2020_3672110}. By this measure (and excluding self-citations to the first and last authors of our current paper), our references contain 11.5\% woman(first)/woman(last), 22.09\% man/woman, 22.36\% woman/man, and 44.05\% man/man. This method is limited in that a) names, pronouns, and social media profiles used to construct the databases may not, in every case, be indicative of gender identity and b) it cannot account for intersex, non-binary, or transgender people.

\vspace{0.3cm}

\noindent Second, we obtained predicted racial/ethnic category of the first and last author of each reference by databases that store the probability of a first and last name being carried by an author of color \cite{ambekar2009name, sood2018predicting}. By this measure (and excluding self-citations), our references contain 8.04\% author of color (first)/author of color(last), 12.89\% white author/author of color, 22.97\% author of color/white author, and 56.11\% white author/white author. This method is limited in that a) names and Florida Voter Data to make the predictions may not be indicative of racial/ethnic identity, and b) it cannot account for Indigenous and mixed-race authors, or those who may face differential biases due to the ambiguous racialization or ethnicization of their names. We look forward to future work that could help us to better understand how to support equitable practices in science.

\section*{Conflict of interest statement}
\noindent The authors declare that they have no known competing financial interests or personal relationships that could have appeared to influence the work reported in this paper.

\section*{Acknowledgments}

\noindent The authors express their gratitude to the collaborators of the CRMBM/CEMEREM UMR CNRS 7339 for their contribution to this research. The authors would like to thank Stanislas Rapacchi for his insightful comments on data processing.

\noindent This research was supported by Université Gustave Eiffel.

%%%%%%%%%%%%%%%%%%%%%%%%%%%%%%
\noindent This version of the article has been accepted for publication, after peer review and is subject to Elsevier's \href{https://www.elsevier.com/about/policies-and-standards/sharing#3-accepted-manuscript}{accepted manuscript terms of use}, but is not the published journal article (PJA) and does not reflect post-acceptance improvements, or any corrections.

\noindent The PJA is available online at: https://doi.org/10.1016/j.clinbiomech.2024.106396.

\noindent This accepted manuscript is licensed under the terms of the \href{https://creativecommons.org/share-your-work/cclicenses/}{CC BY-NC-ND} license.

\section*{Data availability statement}

\noindent The data presented in this study are stored in an open access repository and available on a reasonable request from the corresponding author.

\noindent The DOI associated with intra-abdominal pressure data is: https://doi.org/10.57745/BLKGI0 . 

\noindent The DOI associated with data obtained from segmentation and registration of dynamic axial abdominal MRI is: https://doi.org/10.57745/R5AQRS .

\section*{Declaration of Generative AI and AI-assisted technologies in the writing process}

\noindent During the preparation of this work the authors drafted the article and then used Chat GPT 3.5 in order to improve the readability and language of some paragraphs. After using this tool/service, the authors reviewed and edited the content as needed and take full responsibility for the content of the publication.

\section*{Author contributions}

\noindent \textbf{V.J}: Conceptualisation, Data curation, Writing - original draft

\noindent \textbf{A.J}: Conceptualisation, Investigation, Data curation, Writing - review \& editing

\noindent \textbf{D.B}: Conceptualisation, Writing - review \& editing

\noindent \textbf{A.S}: Conceptualisation, Investigation, Writing - review \& editing

\noindent \textbf{M.G}: Conceptualisation, Writing - review \& editing

\noindent \textbf{C.M}: Conceptualisation, Writing - review \& editing

\noindent \textbf{T.B}: Conceptualisation, Writing - review \& editing

\newpage
% \bibliography{biblio.bib}
% \bibliography{bibexport.bib}
\bibliography{clean_and_diversity_merged.bib}

\end{document}